\documentclass[aps,twocolumn,showpacs,preprintnumbers,amsmath,amssymb,superscriptaddress]{revtex4}
\usepackage{graphicx}
\usepackage{dcolumn}
\usepackage{bm}
\usepackage{color}
\definecolor{slight}{gray}{0.85}
\begin{document}

\title{Tunneling conductance of graphene ferromagnet-insulator-superconductor junctions}
\author{Ya-Fen Hsu}
\affiliation{Department of Physics
and Center for Theoretical Sciences, National Taiwan University,
Taipei 106, Taiwan}
\author{Guang-Yu Guo}
\email{gyguo@phys.ntu.edu.tw} \affiliation{Graduate Institute of Applied
Physics, National Chengchi University, Taipei 116, Taiwan}
\affiliation{Department of Physics
and Center for Theoretical Sciences, National Taiwan University,
Taipei 106, Taiwan}

\date{\today}

\begin{abstract}
We study the transport properties of a graphene
ferromagnet-insulator-superconductor (FIS) junction within the
Blonder-Tinkham-Klapwijk formalism by solving spin-polarized
Dirac-Bogoliubov-de-Gennes equation. 
In particular, we calculate the spin-polarization of tunneling current 
at the I-S interface and invesigate how the exchange splitting of the Dirac fermion
bands influences the characteristic conductance oscillation of the 
graphene junctions. 
We find that the retro and specular Andreev reflections in the graphene 
FIS junction are drastically modified in the presence of exchange interaction and that
the spin-polarization ($P_T$) of tunneling current can be tuned from 
the positive to negative value by bias voltage ($V$).
In the thin-barrier limit, the conductance $G$ of a graphene FIS junction oscillates
as a function of barrier strength $\chi$.
Both the amplitude and phase of the conductance oscillation varies with the exchange energy $E_{ex}$.
For $E_{ex}<E_F$ (Fermi energy), the amplitude of oscillation decreases with $E_{ex}$.
For $E_{ex}^{c}>E_{ex}>E_F$, the amplitude of oscillation increases with $E_{ex}$,
where $E_{ex}^{c}=2E_{F}+U_{0}$ ($U_{0}$ is the applied electrostatic potential
on the superconducting segment of the junction).
For $E_{ex} > E_{ex}^{c}$, the amplitude of oscillation decreases with $E_{ex}$ again.
Interestingly, a universal phase difference of $\pi/2$ in $\chi$ exists between 
the $G-\chi$ curves for $E_{ex}>E_F$ and $E_{ex}<E_F$.
Finally, we find that 
the transitions between retro and specular Andreev reflections occur
at $eV=|E_{F}-E_{ex}|$ and $eV=E_{ex}+E_{F}$, and hence the singular 
behavior of the conductance
near these bias voltages results from the difference in transport properties 
between specular and retro Andreev reflections.
\end{abstract}
\maketitle

\section{Introduction}
Graphene is a single atomic layer of graphite with a two-dimensional (2D) honeycomb lattice structure\cite{Castro}.
The valence and conduction bands of graphene touch each other at six corner points of the 2D hexagonal Brillouin zone.
These six corner points are known as Dirac points and are divided into two inequivalent groups, denoted
by ${\bf K}$ and ${\bf K'}$, respectively.
In 1947, Wallace\cite{Wallace} first predicted that graphene possesses a linear
Dirac-like energy dispersion near each Dirac point.
The linear approximation for energy dispersion is valid even up to 1 eV above
and below the Dirac points.
However, the fabrication of graphene was realized only a few years ago\cite{Geim}.
The linear dispersion relation of graphene has been recently observed in angle-resolved
photoemission spectroscopy (ARPES) experiment\cite{Zhou1}.
Due to the peculiar electronic structure, charge carriers in graphene behave like massless relativistic particles
(called Dirac-like fermions) at low energes.
Recent theoretical works showed that Dirac-like fermions in graphene may result in several novel phenomena
such as anomalous quantum Hall effect\cite{Gusynin} and Klein tunneling\cite{Katsnelson,Klein}.
The anomalous quantum Hall effect\cite{Novoselov,Zhang} and Klein tunneling\cite{Huard,Stander,Andrea} have been observed
recently. In addition, 
the Fermi level of graphene can be tuned by a gate voltage through the electric field effect\cite{Geim}.
It is also possible to create ferromagnetism and superconductivity in graphene via the proximity effect.

Superconductivity is one of the interesting properties of graphene.
In nature, graphene is a zero-gap semiconductor and not a superconductor.
However, superconductivity of graphene may be induced by placing a superconducting electrode on the top of graphene
via the proximity effect\cite{Volkov,Kasumov,Morpurgo,Buitelaar,Herrero}
or coating a sheet of graphene with metallic atoms\cite{Uchoa}.
There have been many studies on superconductivity of graphene.
Beenakker\cite{Beenakker} showed that a normal metal-superconductor (NS)
graphene junction\cite{Rainis} can exhibit unique specular Andreev reflection
which is qualitatively different from usual Andreev reflection in
a conventional NS junction\cite{Blonder}.
Josephson current of a superconductor-normal metal-superconductor (SNS) graphene junction
has also been calculated\cite{Titov,Moghaddam1} and observed\cite{Heersche}.
The observation of Josephson current confirms that superconductivity of graphene
can be induced via the proximity effect.
It was shown that the tunneling conductance of a normal metal-insulator-superconductor (NIS) graphene junction
oscillates with barrier strength(called the conductance oscillation effects)\cite{Bhattacharjee1,Bhattacharjee2,Linder1,Linder2}.
This behavior is analogous to the Klein tunneling.
Notice that the insulator here is not an insulator with an energy gap between the valence and conduction bands.
The notation `` I" in a graphene NIS junction means just a normal segment of graphene with a large electrostatic potential.
Such an insulator could be created by using a gate voltage or chemical doping\cite{Geim}.

Recent theoretical studies showed that intrinsic ferromagnetism of graphene
may exist\cite{Peres1,Peres2,Son}, but it has not been observed so far.
Room-temperature ferromagnetism was observed\cite{Wang},
but it was attribued to the presence of the defects on graphene.
However, proximity-induced ferromagnetism in graphene was recently 
realized experimentally\cite{Tombros}.
Furthermore, a theorectical work showed that the spin injection efficiency for 
Co/Al$_2$O$_3$/graphene could amount to 18\%
and could also be increased up to 31\% by applying a current bias\cite{Jozsa}.
Therefore, it is feasible that a segment of graphene could be ferromagnetic.
Recently, quantum coherence transport in a graphene ferromagnet-superconductor
(FS) junction was investigated theoretically\cite{Zareyan,Asano,Qingyun}.
These studies suggested that the conductance of graphene could increase with
the exchange energy $E_{ex}$ when the $E_{ex}$ is larger than the Fermi energy $E_F$.
This finding is dramatically different from conventional FS 
junctions\cite{Golubov,Chalsani}
and strongly contradicts our intuition.
As stated above, the interplay between superconductivity and ferromagnetism of graphene is interesting.
The tunneling behavior of Dirac-like quasipariticles, the analogous Klein tunneling, is also remarkable.

In this paper, therefore, we study the transport properties of a graphene
ferromagnet-insulator-superconductor (FIS) junction
by using Blonder-Tinkham-Klapwijk (BTK) formalism\cite{Blonder}.
The previous studies on the transport properties of graphene F/S juctions
focus on the effect of the exchange energy on zero-bias conductance\cite{Zareyan} 
and conductance spectra\cite{Asano}.
In contrast, here we calculate the spin-polarization of tunneling current at the
I-S interface and also invesigate how the exchange splitting of the Dirac fermion bands
influences the characteristic conductance oscillation of the graphene junctions. 
In particular, we find that the spin polarization of a graphene ferromagnetic segment
is not always positively correlated with the exchange energy and that the tunneling
spin polarization could be negative.
In addition, we calculate the conductance
in the thin-barrier limit, where thickness $d\rightarrow0$ and the barrier potential $V_0\rightarrow\infty$.
In this limit, the conductance ($G$) of a FIS junction oscillates as a function of the barrier strength $\chi$
and the amplitude of oscillation of conductance varies with $E_{ex}$ and is minimal at $E_{ex}=E_{F}$.
The $G-\chi$ curve for $E_{ex}>E_F$ is phase shifted by $\pi/2$ in barrier strength $\chi$,
compared with the curve for $E_{ex}<E_F$.

\begin{figure}
\includegraphics[width=8cm]{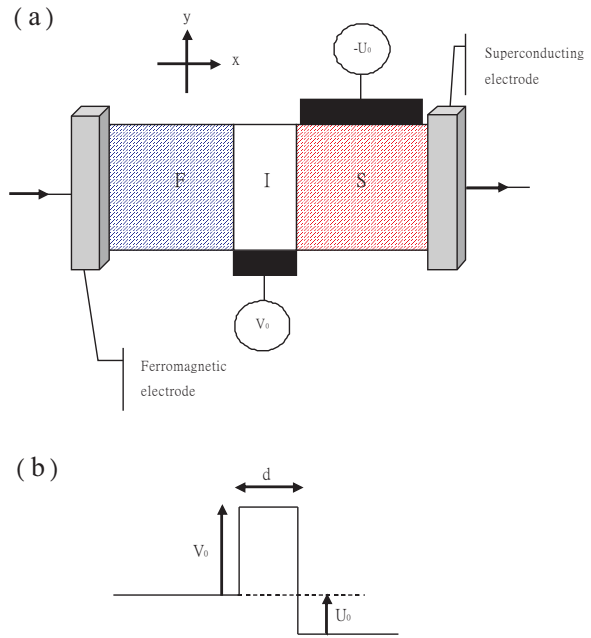}
\caption{(Color online) (a) A schematic plot of a graphene FIS junction.
Here, F, I, and S denote the ferromagnetic, insulating and superconducting regions, respectively.
The F and S regions could be manufactured by contacting graphene with
a ferromagnetic electrode and a superconducting electrode, respectively.
The I region could be created by an external gate voltage $V_0$.
An additional gate voltage $U_0$ may be applied on the S region.
(b) A schematic plot of the electrostatic potential $U({\bf r})$ in the FIS junction of (a).}
\end{figure}

\section{theoretical method and analytical calculations}
\subsection{Theoretical model}
Here we consider a graphene FIS junction, as shown in Fig. 1(a),
with region F in $x\leq-d$, region I in $-d{\leq}x\leq0$ and region S in $x\geq0$.
To tackle a problem concerning superconductivity and relativity,
we should consider the Dirac-Bogoliubov-de-Gennes (DBdG) equation. The DBdG equation
for the graphene nonmagnetic metal-superconductor (NS) interface was derivied in 
Ref. \cite{Beenakker}
by using a time-reversal operator $\mathcal{T}=(\sigma_{z}\otimes\tau_{x})C$\cite{Suzuura}.
Here $C$ is the complex-conjugation operator.
$\sigma_{j}$ and $\tau_{j}$ are the $2\times2$ Pauli matrices acting on sublattices
(i.e. pesdospins) and valleys, respectively.
The final form of the DBdG equation in \cite{Beenakker} was decomposed into two 
decoupled four-dimensional sets
\begin{equation}
\left(\begin{array}[c]{cc}
   H_{\pm}-E_{F}\hat{1} & \Delta\hat{1} \\
   \Delta^\ast\hat{1} & -(H_{\pm}-E_{F}\hat{1})
  \end{array}\right)\left(\begin{array}[c]{cc}
   u^{\pm}\\\nu^{\mp}
  \end{array}\right)=\varepsilon\left(\begin{array}[c]{cc}
   u^{\pm}\\\nu^{\mp}
  \end{array}\right),
\end{equation}
where $\Delta$ denotes the pair potential in the superconducting segment
and $\hat{1}$ is the $2\times2$ indentity matrix.
$H_{\pm}$ is the single particle Hamiltonian with indices $\pm$ labeling
the two valleys of the band structure
at ${\bf K}$ and ${\bf K'}$, respectively.
The single particle Hamiltonian $H_{\pm}$ is given by\cite{Castro}
\begin{equation}
H_{\pm}=-{\it i}\hbar\upsilon_F(\sigma_x\partial_x\pm\sigma_y\partial_y)+U,
\end{equation}
where $\upsilon_F$ represents the Fermi velocity with
$\upsilon_F\backsimeq10^6 m/s$\cite{Castro}.
In Eq. (1), $u$ and $\nu$ denote the wave functions of the electron-like 
and hole-like excitations, respectively.
It is obvious that both $u$ and $\nu$ are composed of two components which
represent sublattices A and B in grephene, respectively.
Note that the solutions of Eq. (1) must satisfy the condition:
$(\nu^{-}\mbox{ }\nu^{+})^{T}=\mathcal{T}(u^{+}\mbox{ }u^{-})^{T}$, where 
$(u^{+}\mbox{ }\nu^{-})^{T}$ and $(u^{-}\mbox{ }\nu^{+})^{T}$ are, respectively,
the solutions of Eq. (1) with $H_{+}$ and $H_{-}$.
In other words, $\nu^{-}=\sigma_{z}Cu^{-}$ and $\nu^{+}=\sigma_{z}Cu^{+}$.
As can be seen from Eq. (1), the electron-like part of wavefunction of ${\bf K}$ 
valley $u^{+}$ is coupled with $v^{-}$,
which is related to the electron-like part of wavefunction of ${\bf K'}$ 
valley $u^{-}$, by the energy gap $\Delta$.
A similar relation exsists between $u^{-}$ and $v^{+}$.
This indicates that the two valleys ${\bf K}$ and ${\bf K'}$ are coupled together
by $\Delta$ for a graphene superconducting segment.

To take ferromagnetism into consideration, we regard the ferromagnetic segment
as a free-electron-like Stoner ferromagnet\cite{Jong}.
Its spin-up band is shifted down by the exchange energy $h$, 
while its spin-down band is shifted up by $h$.
The intrinsic spin-orbit coupling in graphene is negligible\cite{Min}.
Following Ref. \cite{Beenakker} and introducing real spin degrees of freedom, 
we find that the DBdG equation
in the presence of an exchange interaction is given by
\begin{equation*}
\left(\begin{array}[c]{cc}
        H_{\pm}-E_{F}\hat{1}-\rho_{\sigma}h\hat{1} & \Delta\hat{1}\\
        \Delta^\ast\hat{1} & -(H_{\pm}-E_{F}\hat{1}+\rho_{\sigma}h\hat{1})
        \end{array}\right)
  \left(\begin{array}[c]{cc}
        u_{\sigma}^{\pm}\\\nu_{\bar{\sigma}}^{\mp}
        \end{array}\right)
\end{equation*}
\begin{equation}
=\varepsilon\left(\begin{array}[c]{cc}
        u_{\sigma}^{\pm}\\\nu_{\bar{\sigma}}^{\mp}
        \end{array}\right).
\end{equation}
Subscript $\sigma$($\bar{\sigma}$) labels the real spin.
If $\sigma$ denotes up (down) spin, $\bar{\sigma}$ denotes down (up) spin.
$\rho_{\uparrow}$ is 1 and $\rho_{\downarrow}$ is -1.
Note that the indices $\uparrow$, $\downarrow$ refer to 
the two spin subbands, respectively.
In addition to the coupling between the two valleys, Eq. (3) also implies 
that an Andreev reflected hole is formed due to the removal of an electron whose
spin direction is opposite to that of the incident electron.
The hole is an antiparticle of the removed electron.
Therefore, the hole has the same spin direction as the incident electron.
Let us call Eq. (3) the spin-polarized DBdG equation 
and the details of its derivation are given in Appendix A.

We assume that ferromagnetism and superconductivity are uniformly induced
in region F and region S, respectively.
Therefore, $h$, $\Delta$, and $U$ are modeled as step functions,
\begin{equation*}
h({\bf r})=E_{ex}\Theta(-x-d),
\end{equation*}
\begin{equation*}
\Delta({\bf r})=\Delta_{0}e^{{\it i}\phi}\Theta(x),
\end{equation*}
\begin{equation}
U({\bf r})=-U_0\Theta(x)+V_0\Theta(x+d)\Theta(-x),\\
\end{equation}
where $\Theta$ is the Heaviside step function and $\phi$ is the phase of the pair
potential of the superconducting segment.
The potential $U$ is plotted in Fig. 1(b).
However, the DBdG equation is derived within the mean-field approximation\cite{deGennes}.
The mean-field approximation is valid only when the Fermi wave length $\lambda'_{F}$
in region S is much smaller than the superconducting coherence length $\xi$,
where $\xi\propto1/\Delta_0$\cite{Tinkham} and $\lambda'_{F}\propto1/(E_F+U_0)$.
Therefore, our calculations should be done under the condition: $\Delta_0\ll(E_F+U_0)$.
When the order of magnitude of $E_F$ $\mathcal{O}(E_F)\lesssim\mathcal{O}(\Delta_0)$,
we must adopt the regime $U_0\gg\Delta_0$.
We can always satisfy the requirement of the mean field approximation
by adjusting $U_0$ via the gate voltage.

Because of the valley degeneracy, we only need to consider the sets with $H_{+}$ in Eq. (3).
We solve Eq. (3) with $H_{+}$ and obtain the dispersion relations of excitations in region F
\begin{equation*}
\varepsilon_{e\uparrow}  =\hbar\upsilon_F|{\bf k}|-E_{ex}-E_F,
\end{equation*}
\begin{equation*}
\varepsilon^{\pm}_{h\downarrow}=\pm\hbar\upsilon_F|{\bf k}|-E_{ex}+E_F,
\end{equation*}
\begin{equation*}
\varepsilon^{\pm}_{e\downarrow}=\pm\hbar\upsilon_F|{\bf k}|+E_{ex}-E_F,
\end{equation*}
\begin{equation}
\varepsilon^{\pm}_{h\uparrow}=\pm\hbar\upsilon_F|{\bf k}|+E_{ex}+E_F,
\end{equation}
where subscript e (h) indicates electron (hole) excitations.
${+}$ and ${-}$ represent the dispersions of the bands plotted with dotted lines
and solid lines, respectively, in Fig. 2.
The crossing points between the conduction and valence bands are
at $\varepsilon=|E_F-E_{ex}|$ and $\varepsilon=E_F+E_{ex}$.

\begin{figure}
\includegraphics[width=8cm]{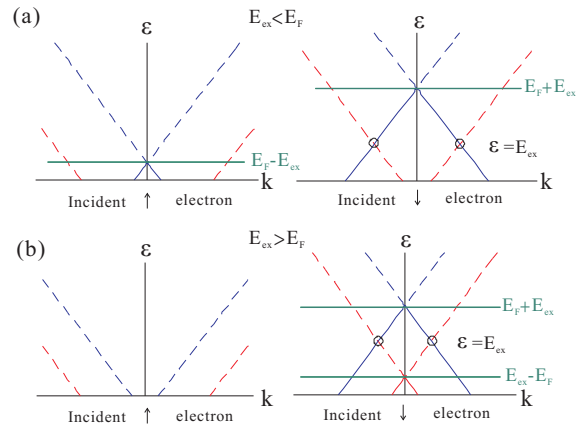}
\caption{(Color online) The excitation spectra in the ferromagnetic
graphene segment in
(a) for $E_{ex}<E_F$ and in (b) for $E_{ex}>E_F$, calculated using
Eq. (5). Red lines indicate electron excitations, while blue lines
indicate hole excitations. Solid and dotted lines
denote the conduction and valence bands, respectively.}
\end{figure}

We also obtain the dispersion relation of quasiparticles in region S \begin{equation}
\varepsilon_{s}=\sqrt{(E_F+U_0\pm\hbar\upsilon_F|{\bf k}|)^2+\Delta_0^2}.
\end{equation}
In addition, the wave functions in region F, region I and region S are also obtained.

\subsection{Calculation of conductance}
We can calculate the Andreev reflection ($a_{\sigma}$) and normal reflection ($b_{\sigma}$) amplitudes
by applying the appropriate boundary conditions to match the wave functions.
The details of the calculation 
are described in Appendix B.
The tunneling conductance is calculated by considering the contributions
of the incident spin-up and spin-down electrons.
The zero-temperature tunneling conductance of the FIS junction is given by
\begin{align*}
&G(eV)=\sum_{\sigma}G_{\sigma}(eV)\int^{\alpha_{\sigma}^c(eV)}_{0}d\alpha_\sigma\cos(\alpha_\sigma)\times\\
&\left(1+|a_\sigma(eV,\alpha_\sigma)|^2\frac{\cos(\alpha'_{\bar{\sigma}})}{\cos(\alpha_{\sigma})}-|b_\sigma(eV,\alpha_\sigma)|^2\right),
\end{align*}
\begin{equation}
G_{\sigma}(eV)=\frac{2e^2}{h}N_{\sigma}(eV),\quad N_{\sigma}(\varepsilon)=\frac{|\varepsilon+E_F+\rho_{\sigma}E_{ex}|W}{\hbar\upsilon_{F}\pi},
\end{equation}
where $G_\sigma$ is the normal-state conductance of the incident spin-${\sigma}$ electron.
$N_{\sigma}$ is the spin-${\sigma}$ density of state (DOS) in a graphene sheet of width $W$.
Finally, we calculate the normalized conductance $G/G_0 $,
where $G_0=\sum_{\sigma}G_{\sigma}=G_\uparrow+G_\downarrow$.
However, a critical angle of incidence $\alpha_{\sigma}^c$ exist in Eq. (7)
because the wave functions would decay as the incident angle
$\alpha_\sigma>\alpha_{\sigma}^c$.
Adopting an approach used in Ref. \cite{Bhattacharjee2},
we do not consider the evanescent wave functions
and hence set the integration limit to the maximum angle of incidence in Eq. (7).
The critical angle can be written as
\begin{equation*}
\alpha_{\sigma}^c(\varepsilon)= \min(\alpha_{\sigma1}^c(\varepsilon),\alpha_{\sigma2}^c(\varepsilon),\alpha_{\sigma3}^c(\varepsilon)),
\end{equation*}
\begin{equation*}
\alpha_{\sigma1}^c(\varepsilon)=\sin^{-1}(\frac{|\varepsilon+\rho_{\sigma}E_{ex}-E_F|}{|\varepsilon+E_F+\rho_{\sigma}E_{ex}|}),
\end{equation*}
\begin{equation*}
\alpha_{\sigma2}^c(\varepsilon)=\sin^{-1}(\frac{|\varepsilon-|E_F-V_0||}{|\varepsilon+E_F+\rho_{\sigma}E_{ex}|}),
\end{equation*}
\begin{equation}
\alpha_{\sigma3}^c(\varepsilon)=\sin^{-1}(\frac{(E_F+U_0)}{|\varepsilon+E_F+\rho_{\sigma}E_{ex}|}).
\end{equation}
When $\alpha_{\sigma}>\alpha_{\sigma1}^c$, the wave function of the Andreev reflected hole decays.
For $\alpha_{\sigma}>\alpha_{\sigma2}^c$, the quasiparticle in the barrier region cannot propagate.
For $\alpha_{\sigma}>\alpha_{\sigma3}^c$, the wave functions of the transmitted quasiparticles are evanescent.
If $\sin(\alpha_{\sigma1}^c)$ is larger than 1 by using Eq. (8),
the Andreev reflected hole do not decay at any angle of incidence.
Similarly, for $\sin(\alpha_{\sigma2}^c)>1$ [$\sin(\alpha_{\sigma3}^c)>1$],
the quasiparticle can always propagate in region I [region S].
Since we do not take any decaying state into account, the critical angle is the minimum among $\alpha_{\sigma1}^c$,  $\alpha_{\sigma2}^c$,
and $\alpha_{\sigma3}^c$.

\subsection{Modified specular and retro-Andreev reflection}
We note that in the Andreev reflection process, the emerging hole is usually reflected
back along the path of the incoming electron\cite{Waldram}.
Nevertheless, this may happen only in the system with $|{\bf k_{e\sigma}}|\thickapprox|{\bf k_{h\bar{\sigma}}}|$.
If $|{\bf k_{e\sigma}}|$ is not equal to $|{\bf k_{h\bar{\sigma}}}|$,
the Andreev reflected hole does not retrace the path of the incident electron anymore\cite{Kashiwaya}.
However, for a graphene superconducting junction, the hole dose not neccessarily
go along the path of the incident electron
even if $|{\bf k_{e\sigma}}|=|{\bf k_{h\bar{\sigma}}}|$.
Two different types of Andreev reflection may happen.
One is the Andreev retroreflection and the other is the specular Andreev reflection\cite{Beenakker}.
In the heavily doped graphene (with $E_{F}\gg\Delta_{0}$), the Andreev retroreflection takes place and
the outgoing hole retraces the path of the incident electron.
In the weakly doped graphene (with $E_{F}\ll\Delta_{0}$), the specular Andreev reflection takes place and
the outgoing hole travels like the normal reflected electron.
The specular Andreev reflection is a unique phenomenon in a graphene superconducting junction.

\begin{figure}
\includegraphics[width=8cm]{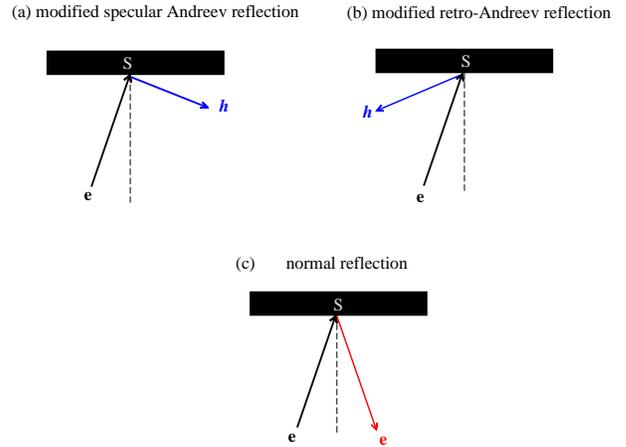}
\caption{(Color online) Schematic plots: (a) the modified specular Andreev reflection,
(b) the modified retro-Andreev reflection, and (c) the normal refletion.}
\end{figure}

For a two or three dimensional system, the boundary conditions would require that
there is no scattering in the transverse directions (i.e., the directions parallel to the interface).
That is to say, the transverse wave vector is conserved in all scattering processes.
For the system investigated here, the transverse direction means the $y$-direction.
The wave vectors of the incident electron and scattered particles in the $y$-direction
are all equal to $q$, as shown in Appendix B.
From this, we deduce that $|{\bf k_{e\sigma}}\sin(\alpha_{\sigma})|=|{\bf k_{h\bar{\sigma}}}\sin(\alpha^{'}_{\bar{\sigma}})|$.
The incident angle $\alpha_{\sigma}$ equates to the Andreev reflected angle $\alpha^{'}_{\bar{\sigma}}$
only when $|{\bf k_{e\sigma}}|=|{\bf k_{h\bar{\sigma}}}|$.
Using ${\bf v=\bigtriangledown_{k}\varepsilon/\hbar}$ and Eq. (5),
where ${\bf v}$ denotes the group velocity, we can determine which direction an excitation goes.
If $(\varepsilon_{e\sigma}+\rho_{\sigma}E_{ex}+E_{F})>0$,
$\varepsilon_{e\sigma}=\hbar\upsilon_F|{\bf k}|-\rho_{\sigma}E_{ex}-E_F$ and
the velocity of the incident electron ${\bf v_{e}}$ is parallel to ${\bf k}$.
Conversely, ${\bf v_{e}}$ is antiparallel to ${\bf k}$.
Similarly, if $(\varepsilon_{h\bar{\sigma}}+\rho_{\sigma}E_{ex}-E_{F})>0$, $\varepsilon_{h\bar{\sigma}}=\hbar\upsilon_F|{\bf k}|-\rho_{\sigma}E_{ex}+E_F$
and the velocity of the reflected hole ${\bf v_{h}}$ is parallel to ${\bf k}$.
Conversely, ${\bf v_{h}}$ is antiparallel to ${\bf k}$.
Therefore, if both $(\varepsilon_{e\sigma}+\rho_{\sigma}E_{ex}+E_{F})$ and
$(\varepsilon_{h\bar{\sigma}}+\rho_{\sigma}E_{ex}-E_{F})$ have the same sign,
$v_{ey}=v_{hy}$. Otherwise, $v_{ey}=-v_{hy}$.
The relation between $\alpha_{\sigma}$ and $\alpha^{'}_{\bar{\sigma}}$ is given by
\begin{equation}
\alpha^{'}_{\bar{\sigma}}=\sin^{-1}(\frac{(\varepsilon+\rho_{\sigma}E_{ex}+E_F)}{(\varepsilon+\rho_{\sigma}E_{ex}-E_F)}\sin(\alpha_{\sigma}))
\end{equation}
Here, the incident angle $\alpha_{\sigma}$ is always positive while $\alpha^{'}_{\bar{\sigma}}$
may be either positive or negative.
If $\alpha^{'}_{\bar{\sigma}}$ is positive, $v_{ey}=v_{hy}$ and the modified specular Andreev reflection happens.
If $\alpha^{'}_{\bar{\sigma}}$ is negative, $v_{ey}=-v_{hy}$ and the modified retro-Andreev reflection occurs.
Here, `` modified" means that $\alpha^{'}_{\bar{\sigma}}\neq\alpha_{\sigma}$.

Kashiwaya {\it et al.} reported that under the effect of the exchange energy the Andreev reflection
angle is not equal to the incident angle any more\cite{Kashiwaya}.
Interestingly, for a doped nonmagnetic graphene superconducting junction
with $\varepsilon$ comparable to $E_F$ and $E_{ex}=0$,
$\alpha^{'}_{\bar{\sigma}}$ is not equal to $\alpha_{\sigma}$ either.
The Fermi energy of a conventinal metal, such as Cu or Al, is in the
order of 1 eV\cite{Ashcroft} and hence
is several thousand times as big as the bias voltage $\varepsilon$.
For conventional nonmagnetic metals, therefore,
$|{\bf k_{e\sigma}}|\thickapprox|{\bf k_{h\bar{\sigma}}}|\thickapprox{k_F}$,
where the Fermi wave vector $k_{F}=\sqrt{2mE_{F}}/\hbar$,
and the incident angle $\alpha_{\sigma}$ is always equal to the
Andreev reflected angle $\alpha^{'}_{\bar{\sigma}}$.
It is interesting that for a graphene superconducting junction,
when $\varepsilon{\approx}E_{F}-\rho_{\sigma}E_{ex}$, the path of the Andreev
reflected hole would severely deviate from the normal line at the small incident angle.
By changing the Fermi energy via a gate voltage, we can reach the condition of
$\varepsilon{\approx}E_{F}-\rho_{\sigma}E_{ex}$ easily.
Fig. 3 illustrates the modified retro-Andreev reflection,
specular Andreev reflection and normal reflection.
From Fig. 3, it can be seen that the normal reflection in a graphene superconducting junction
is not qualitatively different from a conventional superconducting junction.
However, the retro- and specular reflectivity of Andreev reflection
can be drastically broken under the effect of the exchange energy
and the bias for the graphene superconducting junction.

In addition, when either $(\varepsilon+\rho_{\sigma}E_{ex}-E_F)$
or $(\varepsilon+\rho_{\sigma}E_{ex}+E_F)$ changes sign,
there is a trasition between the retro-Andreeve reflection and specular Andreev reflection.
This happens at a Dirac point.
Take Fig. 2(b) for example, the Dirac points of the bands of the spin-down electron and hole excitations are located at $\varepsilon=(E_{ex}-E_{F})$
and $\varepsilon=(E_{ex}+E_{F})$, respectively.
An incident spin-up electron proceeds with the specular Andreev reflection
when $\varepsilon<(E_{ex}-E_{F})$.
As the bias voltage $\varepsilon$ is increased up to the Dirac point of
the band of the spin-down electron($E_{ex}-E_{F}$), the electron would undergo
the retro-Andreev reflection.
Keeping increasing the bias voltage, the electron would proceed with the specular
Andreev reflection again when $\varepsilon\geq(E_{ex}+E_{F})$.
Similarly, a trasition between the retro-Andreeve reflection and specular Andreev
reflection occurs at the Dirac points as a spin-up electron incidents.
In short, for a graphene FIS junction, there are two transition points.
This is quite different from a graphene NIS junction which has only a transition point.

\section{Spin polarization}
\subsection{Spin polarization of DOS}
Measurement of spin polarization is an important subject in spintronics.
There are several ways to measure spin polarization, such as spin-polarized
photoemission spectroscopy\cite{Clauberg},
Meservey-Tedrow spin tunnelling spectroscopy\cite{Tedrow} and
Andreev reflection spectrum\cite{Soulen}.
The BTK formula provide a quite simple method to measure the
spin polarization of a metal by Andreev reflection spectrum.
How to define the spin polarization is also a valuable question
and has been discussed before by Mazin\cite{Mazin}.
The most popular definition of spin polarization in a ferromagnet
is $P=(N_{\uparrow}(E_F)-N_{\downarrow}(E_F))/(N_{\uparrow}(E_F)+N_{\downarrow}(E_F))$,
where $N_{\uparrow}(E_F)$ [$N_{\downarrow}(E_F)$] denotes spin-up [down] DOS at the Fermi level.
For a conventional metal, the Fermi level does not move much when a bias is applied,
and the DOS remains almost constant.
Now consider the effect of the bias on the DOS for graphene.
$N_{\uparrow}$ and $N_{\downarrow}$ are now given in Eq. (7). Therefore,
the spin polarization of DOS can be written as,
\begin{equation}
P=\frac{(\varepsilon+E_{F}+E_{ex})-|\varepsilon+E_{F}-E_{ex}|}{(\varepsilon+E_{F}+E_{ex})+|\varepsilon+E_{F}-E_{ex}|}.
\end{equation}

The spin polarization in a graphene ferromagnetic segment varies with the applied voltage.
When $\varepsilon \leq E_{ex}-E_F$, $P=(\varepsilon+E_F)/E_{ex}$.
Conversely, $P=E_{ex}/(\varepsilon+E_F)$. In the low bias regime ($E_{ex}, E_{F} \gg \varepsilon$),
we find that $P=E_{ex}/E_{F}$
if $E_{ex} \leq E_{F}$ and $P=E_{F}/E_{ex}$ if $E_{ex} > E_F$.
The spin polarization becomes independent of $\varepsilon$.
Furthermore, it is counterintuitive that
the spin polarization is not always positively correlated with $E_{ex}$.
It is clear that we could increase the spin polarization
in graphene by tuning the Fermi energy via a gate voltage.
The Fermi energy in graphene is usually in the order of 0-100 meV.
For the lightly doped graphene, therefore, we only need to apply a gate voltage of a few milli-electron
voltage to change the spin polarization in graphene significantly.
Furthermore, for $E_{F}=E_{ex}\gg\varepsilon$, graphene is a half metal with $P=1$.
The exchage energy is fixed by the proximity effect.
Hopefully, this interesting prediction of half-metallic graphene will
stimulate future experiments on graphene by tuning the gate voltage to
the condition of $E_{F}=E_{ex}$.

Nonetheless, it is worth mentioning that when $E_{F}$ is in the vicinity 
of the neutrality point (i.e. the Dirac point), the carrier density goes to zero.
The electron-hole puddles can appear at low carrier densities and in the presence 
of disorder.\cite{Hwang,Martin} In this case, local variations of the Fermi 
energy should be taken into account.
It is well known that a perfect graphene sheet is very difficult to obtain 
experimentally\cite{Martin}. Therefore, in experiments, corrugations are one source 
of disorder. Fascinatingly, it has been very recently demonstrated that an ultraflat 
graphene with height variation less than 25 pm can be obtained\cite{Lui},
suggesting that disorder induced behaviors such as electron-hole puddle formation,
of graphene could be avoided.
In a ferromagnetic graphene, the carrier density of spin-down electrons would approach 
zero in the vicinity of $E_{ex}$ at $E_{F}=E_{ex}$.
However, unlike in nonmagnetic graphenes, 
the Dirac points of different spin species in a ferromagnetic graphene are not 
located at the same position. Therefore, the total carrier density is not 
necessarily close to zero as $E_{F}$ approaches $E_{ex}$,
and our calculations would be still valid in the sufficiently clean (little dirty) regime.

\begin{figure}
\includegraphics[width=7cm]{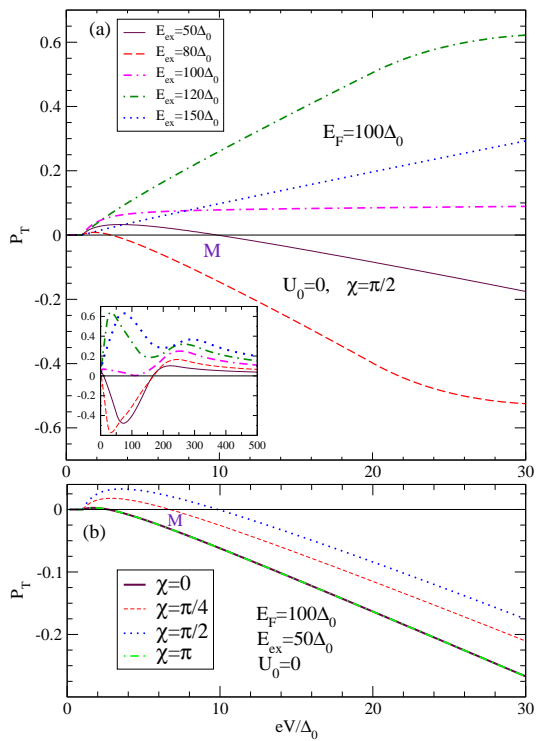}
\caption{(Color online) (a)The tunneling spin polarization
for several values of the exchange energy
$E_{ex}$ with $E_F=100\Delta_0$, $U_0=0$ and $\chi=\pi/2$.
The insert: the $P_{T}$-$eV$ curve with the same parameters
as (a) and $eV$ in the range of 0-500$\Delta_{0}$.
For $E_{ex}<E_F$, the spin polarization could be negative.
(b)The effect of barrier on the curve of the tunneling spin polarization versus bias voltage
for $E_{F}, E_{ex}\gg\Delta_{0}$
with $E_F=100\Delta_0$, $E_{ex}=50\Delta_0$, and $U_0=0$.
}
\end{figure}

\subsection{Tunneling spin polarization}
The spin polarization of tunneling current for a graphene FIS junction is defined as
\begin{equation}
P_{T}=\frac{(I_{\uparrow}-I_{\downarrow})}{(I_{\uparrow}+I_{\downarrow})},
\end{equation}
where $I_{\sigma}$ is the spin-$\sigma$ current which injects into 
the superconductor at the I/S interface.
The detailed calculation of $P_{T}$ is displayed in Appendix C.
It should be emphasized that the magnitude of the tunneling spin polarization 
$P_{T}$ here represents the spin injection efficiency of a graphene FIS junction 
at the I/S interface.
Due to spin diffusion\cite{Ziese}, the spin current in the junction 
does not remain constant, and indeed, deep inside the superconductor, $P_{T}$ of 
the tunneling current would go to zero.

We show the spin polarization of tunneling current as a function of bias voltage 
in Fig. 4 (a).
Here, we pay more attention on the bias in the range of $0-30\Delta_{0}$
because the Andreev reflection spectrum is usually measured 
at $eV\approx0-10$ meV\cite{Soulen, Chiang}.
The enegy gap $\Delta_{0}$ is around $0.5$ meV\cite{Bhattacharjee2}.
Nonetheless, $P_{T}$-$eV$ curves with $eV$ in the range of $0-500$ meV are also plotted
in the insert in Fig. 4(a).
From Fig. 4(a), we find the tunneling spin polarization $P_{T}$ is always zero at subgap bias for any $E_{ex}$.
This is because only supercurrent can flow through the interface for $E<\Delta_{0}$ 
and singlet supercurrent carries no spin polarization.
For $E_{ex}<E_F$, excluding $E_{ex} \approx E_{F}$, $P_{T}$ slightly increases
and then decreases with increasing $eV$ up to around $eV=E_{F}-E_{ex}$.
Taking the curve with $E_{ex}=50\Delta_{0}$ as example, at around $eV=E_{F}-E_{ex}$, the tunneling spin polarization reaches its minimum.
Futhermore, after $P_{T}$ reaches the minimum at around $eV=E_{F}-E_{ex}$, $P_{T}$ is raised by $eV$ again.
Untill around $eV=E_{F}+E_{ex}$, a local maximum appears, and the $P_{T}$ then decreases
monotonically as $eV$ increases.
Finally, $P_{T}$ approaches $0^{+}$ at high bias.
Interestingly, $P_{T}$ would become negative when the bias
is increased up to some critical value $M$ (Fig. 4a),
which is the transition point between the positive and negative tunneling 
spin polarization.
However, for $E_{ex}>E_{F}$, excluding $E_{ex} \approx E_{F}$, $P_{T}$
increases up to its maximum and then falls. The maximum happens at 
around $eV=E_{ex}-E_{F}$.
After $P_{T}$ falls to some degree, it would increase with $eV$ again untill $eV$ is about $eV=E_{ex}+E_{F}$.
A local maximum occurs at about $eV=E_{ex}+E_{F}$.
After $P_{T}$ reaches the local maximum at around $eV=E_{ex}+E_{F}$, 
it decreases monotonically  with increasing $eV$.
It is clear from Fig. 4(a) that unlike the case with $E_{ex}<E_{F}$, the tunneling 
spin polarization for $E_{ex}>E_{F}$ is always positive.

In short, the magnitude of tunneling spin polarization $|P_{T}|$ reaches its maximum
at around $eV=|E_{F}-E_{ex}|$ and a local maximum at around $eV=E_{F}+E_{ex}$ 
for any $E_{ex}$ which is not close to $E_{F}$
and the graphene FIS junction can exhibit the unique negative-value tunneling 
spin polarization if $E_{ex}<E_{F}$.
The negative tunneling spin polarization implies that it is harder for 
majority carriers (spin-up electrons) to tunnel through the interface 
than for minority carriers (spin-down electrons).
$eV=|E_{F}-E_{ex}|$ and $eV=E_{F}+E_{ex}$ are the Dirac points of the spin-down 
electron and hole bands, as shown in Fig. 2.
The Andreev reflection is suppressed at the Dirac points of the hole bands.
No spin-polarized current can flow through the interface via the Andreev 
reflection process. Therefore, near the Dirac points of the hole band, 
spin-polarized current is easier to be carried over
and hence the tunneling spin polarization can be large when $eV$ is 
located at the Dirac points of the hole bands.
Near the Dirac points of the spin-down electrons, the spin-down DOS 
is very small and therefore spin-up current predominates.
In this case, the spin-down current is small and hence the tunneling 
spin polarization would be also large.
We find that the bias voltage can be used to tune the tunneling spin polarization.
Taking the case with $E_{ex}=120\Delta_{0}$ as an example, $P_{T}$ can 
vary from $0\%$ to $62\%$ (i.e., $0-0.62$) in the range of $0-30$ meV.
The effect of barrier on the magnitude of tunneling spin polarization is 
shown in Fig. 4(b). The magnitude of the spin polarization change 
under the effect of barrier strength can be, e.g., $9\%$ ($0.09$) 
at $eV=30\Delta_{0}$ for the case with $E_F=100\Delta_0$, $E_{ex}=50\Delta_0$, and $U_0=0$.
Furthermore, we notice that $\chi$ could move the position of the transition point $M$
(Fig. 4b).

\begin{figure}
\includegraphics[width=7cm]{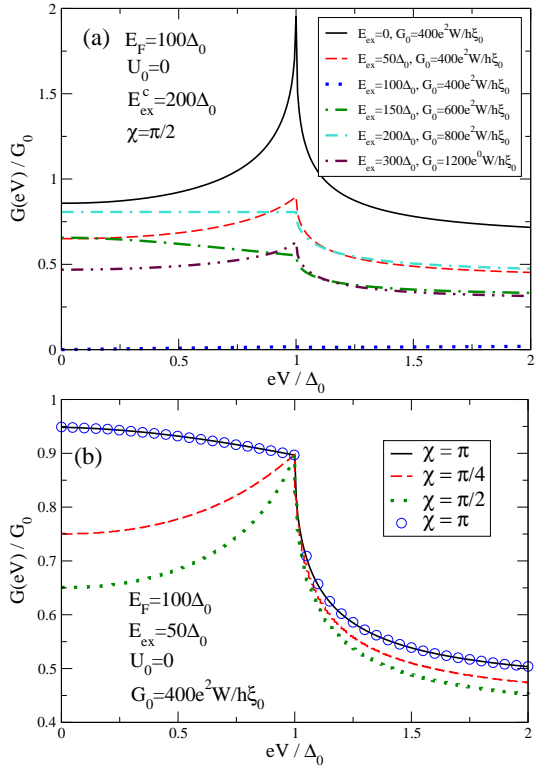}
\caption{(Color online) (a) The conductance spectra for several values of the exchange energy
$E_{ex}$ with $E_F=100\Delta_0$, $U_0=0$ and $\chi=\pi/2$.
(b) The conductance spectra for several values of barrier strength $\chi$
with $E_F=100\Delta_0$, $E_{ex}=50\Delta_0$, and $U_0=0$.
The curves for $\chi=0$ and $\chi=\pi$ are identical.
Note that $G_0$ is not always constant.
For $E_{F}$, $E_{ex}{\gg}eV$, $G_0$ is independent of $eV$.
When $E_{ex}\leq E_{F}$, $G_{0}=4(E_{F}/\Delta_{0})(e^{2}W/h\xi_{0})$, where $\xi_{0}={\pi}{\hbar}v_{F}/\Delta_{0}$.
Conversely, $G_{0}=4(E_{ex}/\Delta_{0})(e^{2}W/h\xi_{0})$.}
\end{figure}

\section{Numerical Results}
\subsection{Conductance in the large Fermi energy limit}
In the limit of a thin barrier, where $d\rightarrow0$ and $V_0\rightarrow\infty$,
we use Eq. (7) and Eq. (A9) to calculate the conductance.
First, we consider the case of $E_F\gg\Delta_0$.
Because $V_0\rightarrow\infty$, $\alpha_{\sigma2}^c$ must be larger than $\alpha_{\sigma1}^c$
and $\alpha_{\sigma3}^c$, where $\alpha_{\sigma1}^c$, $\alpha_{\sigma2}^c$ and  $\alpha_{\sigma3}^c$ are defined in Sec. II.
Here, we are interested in the conductance for a small bias voltage ($0<\varepsilon<2\Delta_0$).
Using Eq. (8), we find that $\alpha_{\downarrow}^c$ is always equal to $90^{0}$ in the low bias regime.
Therefore, we only need to discuss $\alpha_{\uparrow}^{c}$.
If $E_{ex}<2E_F+U_0$, $\alpha_{\uparrow1}^c<\alpha_{\uparrow3}^c$ and $\alpha_{\uparrow}^c=\alpha_{\uparrow1}^c$.
Conversely, $\alpha_{\uparrow1}^c>\alpha_{\uparrow3}^c$ and $\alpha_{\uparrow}^c=\alpha_{\uparrow3}^c$.
We define the critical exchange energy $E_{ex}^{c}=(2E_F+U_0)$.
Here, we adopt $E_F=100\Delta_0\gg\Delta_0$ and $U_{0}=0$.
The critical exchange energy is equal to $200\Delta_{0}$.

The results with $E_F\gg\Delta_0$ are displayed in Figs. 5-7.
As a check of the validity of our calculations, we emphasize that the results of
a graphene NIS junction reported in Ref. \onlinecite{Linder2} can be reproduced by setting $E_{ex}=0$.
It can be seen by comparing the curve for $E_{ex}=0$ in Fig. 5(a) with the results of Ref. \cite{Linder2}.
We can also reproduce the previous results of a graphene FS junction
from Ref. \cite{Zareyan} and Ref. \cite{Asano} by setting $\chi=0$.
This also indicates that our formalism is more general.
From Fig. 5(a), we find that for the large $E_F$ ($E_F\gg\Delta_0$),
the conductance of a graphene FIS junction increases with $E_ {ex} $ for $E_{ex}^{c}>E_ {ex}>E_F$.
Conversely, for $E_{ex}<E_{F}$, the conductance would decrease with increasing $E_{ex}$.
However, for $E_{ex} > E_{ex}^{c}$, the conductance decreases with increasing $E_{ex}$ again.
In principle, the spin polarization ($P$) of DOS would suppress the Andreev reflection\cite{Jong}.
However, when $E_{F}\gg\varepsilon$ (i.e., $\varepsilon\rightarrow{0}$), Eq. (10) indicates that
$P$ increases with increasing $E_{ex}$ for $E_{ex}<E_{F}$
and decreases with increasing $E_{ex}$ for $E_{ex}>E_{F}$ if $E_{F}$ is fixed.
Therefore, when $E_{ex}<E_{ex}^{c}$, the conductance would show the behavior as seen in Fig. 5(a).
At $E_{ex}=E_{F}$, $P$ reaches its maximum value of 1.0, and hence the conductance becomes very small.
However, for $E_{ex}>E_{ex}^{c}$, although $E_{ex}$ does not lower the probability of Andreev reflection,
it would obstruct the propagation of quasiparticles in the superconducting segment.
As $E_{ex}>E_{ex}^c$, $\alpha_{\uparrow}^c$ is equal to $\alpha_{\uparrow3}^c$.
$\alpha_{\uparrow3}^c$ decreases with increasing $E_{ex}$ and hence the
transimission of quasiparticles in region S falls as $E_{ex}$ raises.
Therefore, $E_{ex}$ reduces the conductance even if $P$ is enhanced by $E_{ex}$.

\subsection{Conductance vs barrier strength}
From Fig. 5(b), we find the conductance still oscillates as a function
of $\chi$ with a period $\pi$ like a graphene NIS junction under the effect of exchange energy.
This can also be understood from Eq. (A9).
Both the normal and Andreev reflection amplitudes ($a_\sigma$ and $b_\sigma$) are oscillatory
functions of $\chi$ with a period $\pi$.
Therefore, the conductance $G(eV)$ oscillates as a function of $\chi$ with a period $\pi$.
However, this unique oscillation originates from the relativity of quasiparticles in graphene.
Here the relativity means that quasiparticles in graphene possess a linear energy dispersion.
Our calculation shows that the relativity is not destroyed by exchange splitting.
Dirac-like quasiparticles can transmit through a high barrier, in contrast to nonrelativistic particles.
The motion of non-relativistic particles is described by the Schr\"{o}dinger equation.
Adopting the step-barrier model used in our calculation (see Fig. 1(b)), we find
the wave functions of nonrelativistic particles always decay in the barrier region for $E<V_0$.
The wave functions of nonrelativistic particles are traveling waves and resonance
scattering\cite{Sakurai} may happen only for $E>V_0$.
However, for Dirac-like quasiparticles, the wave functions are traveling waves and
resonance scattering may happen even if $E<V_0$.
As the wave functions are traveling waves, the transmission coefficient must be an oscillatory function of $k_bd$, where $k_b$
is the wave vector of particles in the barrier region.
We let $V_0\rightarrow\infty$ and obtain $k_bd\rightarrow\chi$.
Therefore, the conductance oscillates with $\chi$ even if $V_0\rightarrow\infty$.

\begin{figure}
\includegraphics[width=8cm]{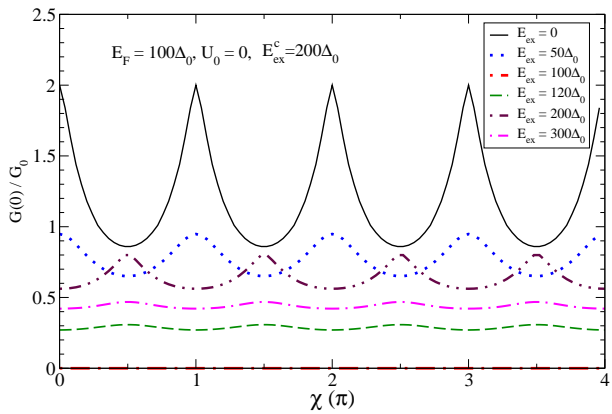}
\caption{(Color online) The zero-bias conductance as a function of barrier strength $\chi$
with $E_F=100\Delta_0$, $U_0=0$, and $E_{ex}=0, 50\Delta_0, 100\Delta_0, 200\Delta_0, 300\Delta_0$.}
\end{figure}

\begin{figure}
\includegraphics[width=7.5cm]{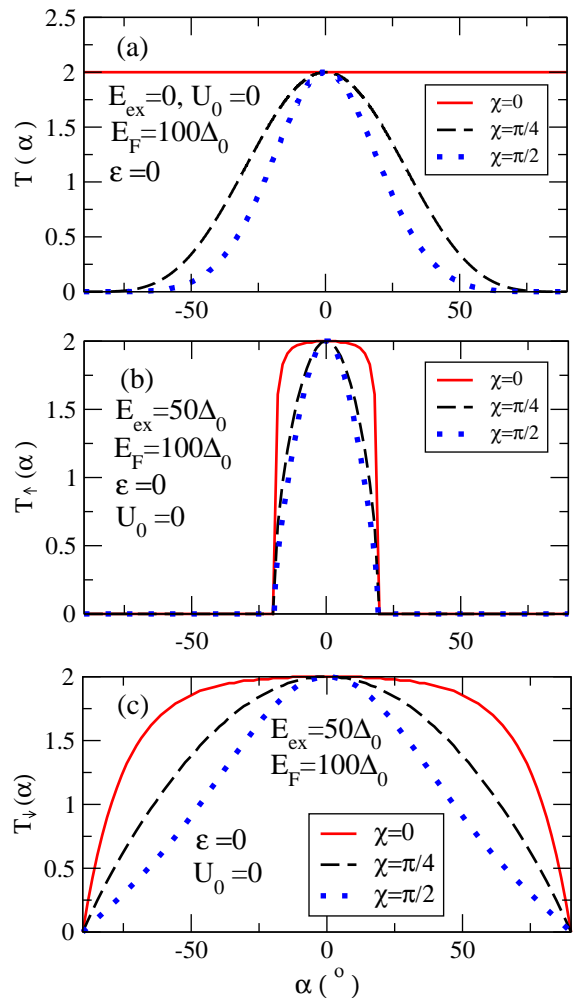}
\caption{(Color online) (a) The transmission coefficient as a function of incident angle $\alpha$
in a NIS junction with $\varepsilon=0$, $E_F=100\Delta_0$, and $U_0=0$ for $\chi = 0, \pi/4, \pi/2$, respectively.
(b) The spin-up transmission coefficientin $T_{\uparrow}$ in a FIS junction
with $\varepsilon=0$, $E_{ex}=50\Delta_0$,
$E_F=100\Delta_0$, and $U_0=0$, where $T_{\uparrow}$ refers to $(1+A_{\uparrow}-B_{\uparrow})$.
(c) The spin-down transmission coefficient $T_{\downarrow}$ in a FIS junction
with $\varepsilon=0$, $E_{ex}=50\Delta_0$,
$E_F=100\Delta_0$, and $U_0=0$, where $T_{\downarrow}$ refers to $(1+A_{\downarrow}-B_{\downarrow})$.}
\end{figure}

Next, let us consider the zero-bias conductance as a function of
barrier strength $\chi$, plotted in Fig. 6.
The zero-bias conductance oscillates with $\chi$ over a period of $\pi$.
The oscillatory behavior results from the Klein tunneling in a superconducting junction.
The Klein tunneling has two remarkable characteristics: i) the transimission
is not always suppressed by a barrier and
ii) for normal incidence, the barrier is perfectly transparent.
Both of the two characteristics have been observed\cite{Huard,Andrea}.
In addition, we find that for $E_{ex}<E_F$, $G(0)$ reaches its maximum at $\chi=n\pi$ where
$n$ is an integer (i.e., $n=0, 1, 2,...$).
In contrast, for $E_{ex}>E_F$, $G(0)$ reaches its maximum at $\chi=(n+1/2)\pi$,
i.e., $G(0)$ is phase shifted by $\pi/2$ in $\chi$, compared with that for $E_{ex}<E_F$.
Due to the phase shift of $\pi/2$, in the reange of $\chi=0$-$\pi/2$, for $E_{ex}>E_F$,
the conductance increases steadily with increasing $\chi$, while for $E_{ex}<E_F$,
the conductance drops sharply as we increase the barrier potential.
For $E_{ex}<E_{F}$, the increase of conductance with increasing $\chi$ can not happen at $\chi<\pi/2$.
Therefore, we would need a very large barrier potential to let $\chi>\pi/2$ in order to
observe the first chacteristic of the Klein tunneling for $E_{ex}<E_{F}$.
This implies that obseving the first chacteristic of the Klein
tunneling for $E_{ex}>E_{F}$ is easier.
We also find that $E_{ex}$ can affect the amplitude of oscillation.
For $E_{ex}<E_F$, the amplitude of oscillation decreases as $E_{ex}$ increases.
The amplitude of oscillation become zero as $E_{ex}=E_F$.
However, for $E_{ex}^{c}>E_{ex}>E_F$, the amplitude of oscillation increases as $E_{ex}$ increases.
For $E_{ex} > E_{ex}^{c}$, the amplitude of oscillation decreases with increasing $E_{ex}$ again.
We could also alter the amplitude of oscillation by applying $U_{0}$\cite{Linder2}.
However, unlike $E_{ex}$, $U_{0}$ would only decrease the amplitude of oscillation.
As $E_{ex}^{c}>E_{ex}>E_{F}$, the exchange energy would enhance the amplitude of oscillation.
In addition, the phase shift of $\pi/2$ is not seen by adjusting $U_{0}$.
These interesting features of $G(0)$-$\chi$ curve can be understood as follows.

\subsection{Transmission vs incident angle}
From Fig. 7(a), it can be seen that the transmission coefficient for
a NIS junction is insensitive to $\chi$ as the incident angle becomes very small.
Here, we regard the transport at very small angle as the nearly normal incidence.
Also note that the transmission coefficient $T$ refers to $(1+A-B)$ for a superconducting junction,
where $A$ and $B$ denote the Andreev and normal reflection coefficients, respectively.
However, $A=|a|^{2}\cos(\alpha')/\cos(\alpha)$ and $B=|b|^2$ for a NIS junction,
where $\alpha'$ is the Andreev reflection angle and $\alpha$ is the incident angle.
The transmission coefficient for a normal metal-insulator-normal metal
junction (NIN) also has the same feature\cite{Castro, Katsnelson}.
Fig. 7(b) and (c) show that the transmission coefficient also depends
on the spin of incident electrons for the FIS junction.
This is different from a NIS junction.
In addition, comparing Fig. 7(b) with Fig. 7(a), we find the critical
angle changes under the effect of $E_{ex}$.
It is clear from Fig. 7(b) and Fig. 7(c) that the insensitivity of the
transmission coefficient to $\chi$ for nearly normal incidence exists
also in the FIS junction, and this is not affected by the exchange spliting.
We conclude that the barrier potential has little effect on the transport
of Dirac-like quasiparticles for the nearly normal incidence.

In addition, Fig. 7(b) shows that the transimission sharply drops to zero as $\alpha$
is close to the critical angle regardless of $\chi$.
The transmission is independent of $\chi$ if the incident angle equates to $0$ or $\alpha_{c}$.
The barrier strength $\chi$ could affect the transimission only in the range of $\alpha=0-\alpha_{c}$.
Furthermore, we find that $T(\chi=0)$ is larger than $T(\chi=\pi/2)$ at any incident angle for $E_{ex}<E_{F}$.
Conversely, for $E_{ex}>E_{F}$, $T(\chi)$ is always less than $T(\chi=\pi/2)$.
For brevity, we do not represent the transmission of the case with $E_{ex}>E_{F}$.
This indicates that the contribution of any incident angle would increases
the amplitude of oscillation ($G_{max}-G_{min}$).
In short, the increase of critical angle could enhance the amplitude of oscillation.

However, as stated in Sec. II. A, we do not consider the evanescent wave
functions when we calculate the conductance of the FIS junction.
As stated above, in the low bias regime, for $E_{ex}<E_{ex}^{c}$,
$\alpha_{\uparrow}^c=\alpha_{\uparrow1}^c$.
From Eq. (8), we see that if $E_{ex}<E_{F}$, $\alpha_{\uparrow1}^c$
decreases with increasing $E_{ex}$ and
conversely, $\alpha_{\uparrow1}^c$ increases with increasing $E_{ex}$.
However, for $E_{ex}^{c}>E_{ex}>E$, $\alpha_{\uparrow}^c=\alpha_{\uparrow3}^c$.
$E_{ex}$ would reduce $\alpha_{\uparrow3}^c$.
If the critical angle is smaller, the conductance is less sensitive to $\chi$
and the amplitude of oscillation would decrease.
Therefore, in Fig. 6, the $G(0)$-$\chi$ curve would display such behavior
(as stated above) under the effect of $E_{ex}$.

Now, we discuss the case of $E_{ex}=E_{F}=100\Delta_{0}$ because
it is interesting that the spin polarization is 1.0.
The critical angle $\alpha_{\uparrow}^c$ equates to $\sin^{-1}(|\varepsilon+E_{ex}-E_F|/|\varepsilon+E_F+E_{ex}|)$.
For the small bias voltage ($\varepsilon\ll{E_F}$),
$\alpha^c=\sin^{-1}(|\varepsilon+E_{ex}-E_F|/|\varepsilon+E_F+E_{ex}|\approx0$.
The Dirac fermion only proceeds with nearly normal incidence and hence the transmission
coefficient is almost independent of $\chi$.
Moreover, the conductance for $E_{ex}=E_F$ is also almost independent of $\chi$.

\begin{figure}
\includegraphics[width=7cm]{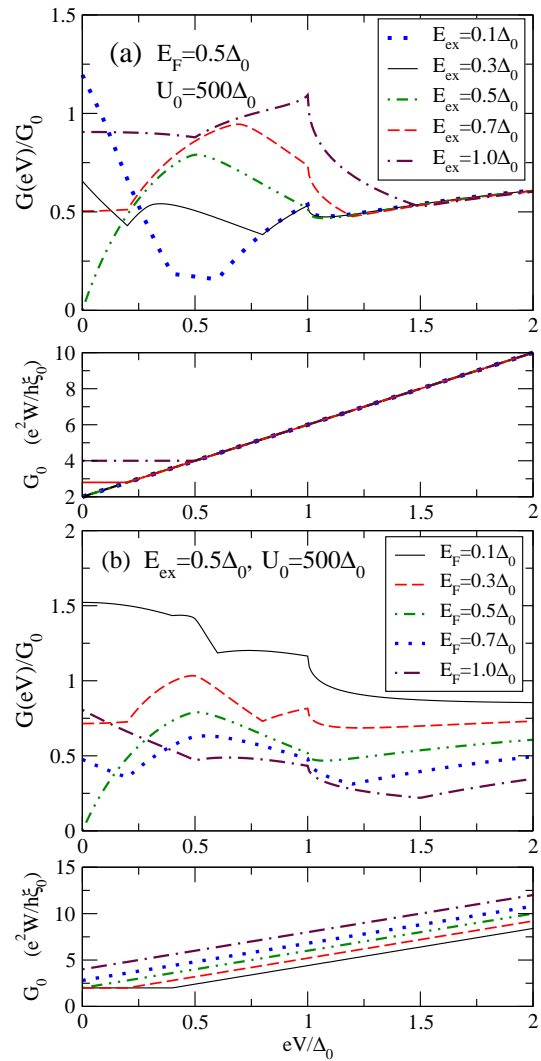}
\caption{(Color online) The conductance spectra (a) for several values of $E_{ex}$ with $E_F=0.5\Delta_0$ and $U_0=500\Delta_0$
and (b) for several values of $E_{F}$ with $E_{ex}=0.5\Delta_0$ and $U_0=500\Delta_0$.
The curves for $\chi=0$ and $\chi=\pi$ are identical.
Note that $G_0$ is not a constant and $\xi={\pi}{\hbar}v_{F}/\Delta_{0}$.
As $\varepsilon\leq(E_{ex}-E_{F})$, $G_{0}=4(E_{ex}/\Delta_{0})(e^{2}W/h\xi_{0})$
and as $\varepsilon>(E_{ex}-E_{F})$, $G_{0}=4((\varepsilon+E_{F})/\Delta_{0})(e^{2}W/h\xi_{0})$, where $\xi_{0}={\pi}{\hbar}v_{F}/\Delta_{0}$.}
\end{figure}

\subsection{Conductance in the small Fermi energy limit}
Now let us discuss the case with comparable $E_{F}$,$E_{ex}$ and $\Delta_0$.
Here, we should take a large $U_0$ to fulfil the requirement of the mean field approximation.
For $U_0>>E_{F}, \Delta_0$, we let $\sin(\gamma)\rightarrow0$ in Eq. (A9) and Eq. (A10).
We find that $a_\sigma$ and $b_\sigma$ are independent of $\chi$.
The conductance also becomes independent of $\chi$.
From Fig. 8, we find that the conductance spectra have singular points
at $eV=|E_{F}-E_{ex}|$, $E_{ex}+E_{F}$, and $\Delta_0$.
Among them, the singular point at $eV =\Delta_0$ can be found in the conductance
spectra of all kinds of superconducting junctions.
Therefore, we do not discuss it any more and focus on the other singular points instead.
The singular points at $eV=|E_{F}-E_{ex}|$, $E_{ex}+E_{F}$ are just the crossing points
of the conduction and valence bands, i.e., the Dirac points shown in Fig. 2.
The appearance of these singular points indicates that the transport properties of quasiparticles
drastically change when they go through these Dirac points.
In fact, at these Dirac points, the transitions between the specular Andreev
reflection and Andreev retroreflection happen, as mentioned in Sec. II. C.
The transitions result in the singular behavior of the conductance near
these two singular points.
The positions of the singular points are related to the exhange energy and Fermi energy.
This indicates that by observing the Andreev reflection spectrum
and labeling the positions of the singular points,
we can measure the exchange energy and Fermi energy of graphene.

\section{conclusions}
In conclusion, we have investigated the transport properties of a graphene
FIS junction within the BTK formalism.
Unlike the previous works on grephene FS junctions\cite{Zareyan, Asano}, 
among other things, we study the spin-polarization of tunneling current 
at the I-S interface and also invesigate how the exchange energy 
influences the characteristic conductance oscillation behavior in graphene junctions.
We find that the spin polarization $P$ of DOS in a graphene ferromagnetic 
segment is not always positively correlated with the exchange energy 
and can be tuned all the way up to full polarization $P = 1$ (i.e., half-metallic
state) with the applied voltage, and that graphene FIS junction
can exhibit the unique negative tunneling spin polarization ($P_T <0$).
In addition, we have calculated the tunneling conductance of the graphene FIS
junction in the thin-barrier limit. We shows in this limit that the tunneling conductance
oscillates as a function of barrier strength ${\chi}$ with a period of ${\pi}$,
and the exchange energy $E_{ex}$ can affect both the amplitude and phase of oscillation.
For $E_{ex}<E_F$, the amplitude of oscillation decreases as $E_{ex}$ increases.
However, for $E_{ex}^{c}>E_{ex}>E_F$, the amplitude of oscillation increases
as $E_{ex}$ increases, where $E_{ex}^{c}=2E_{F}+U_{0}$.
For $E_{ex} > E_{ex}^{c}$, the amplitude of oscillation decreases with $E_{ex}$ again.
The curve for $E_{ex}>E_F$ is phase shifted by $\pi/2$ in $\chi$,
compared with the curve for $E_{ex}<E_F$.
The conductance spectra always have singular points at $eV=|E_{F}-E_{ex}|$, $E_{ex}+E_{F}$ and $\Delta_0$.
This property suggests that one could measure the exchange energy and Fermi energy of graphene
by observing the Andreev reflection spectrum and locating the positions of the singular points.
The unique oscillation originates from the relativity of quasiparticles in graphene.
The wave functions of relativistic quasiparticles with a linear
energy dispersion relation do not decay in the high barrier.
$E_{ex}$ can affect the amplitude of oscillation because the critical
angle $\alpha_\sigma^c$ varies with $E_{ex}$.
The transitions between retroreflection and specular reflection happen
at $eV=|E_{F}-E_{ex}|$ and $eV=E_{ex}+E_{F}$.
The difference in transport properties between specular Andreev reflection
and Andreev retroreflection results in the singular behavior of the conductance
near these singular points.

\section*{ACKNOWLEGEMENTS}
The authors thank Tien-Wei Chiang for valuable discussion.
The authors also acknowledge financial supports from National Science Council and NCTS of Taiwan.

\section*{APPENDIX A: DERIVATION OF SPIN-POLARIZED-DIRAC-BOGOLUBOV-DE-GENNES EQUATION}
The graphene Hamiltonian including both valleys $\bf {K}$ and $\bf {K'}$ is given by
\begin{equation*}
H=\left(\begin{array}[c]{cc}H_{+} & 0\\ 0 & H_{-}\end{array}\right),
\end{equation*}
\begin{equation*}
H_{\pm}=-{\it i}\hbar\upsilon_F(\sigma_x\partial_x\pm\sigma_y\partial_y)+U. \tag{A1}
\end{equation*}
Taking exchange splitting into consideration and introducing real spin degrees of freedom, 
the graphene Hamiltonian can be rewritten as
\begin{equation*}
H'=H\otimes\hat{1}_{2\times2}+h\hat{1}_{4\times4}\otimes{S_{z}}.\tag{A2}
\end{equation*}
$H'$ acts on an eight-dimensional wave function with two sublattices, two valleys  and two 
real spin components,
\begin{equation*}
\Psi=\sum_{\sigma}\varphi^{\sigma}\otimes\zeta_{\sigma},\tag{A3}
\end{equation*}
with
\begin{equation*}
\varphi^{\sigma}=\left(\begin{array}{cccc} \Psi^{\sigma}_{A+} & \Psi^{\sigma}_{B+} & \Psi^{\sigma}_{A-} & \Psi^{\sigma}_{B-}\end{array} \right)^{T},
\end{equation*}
\begin{equation*}
\zeta_{\uparrow}=\left(\begin{array}{cc}1 \\ 0\end{array} \right),\mbox{ }\zeta_{\downarrow}=\left(\begin{array}{cc}1 \\ 0\end{array} \right).\tag{A4}
\end{equation*}
The motion of the quasiparticles in a superconductor should be described by the Bogolubov-de Gennes equation 
\begin{equation*}
\left(\begin{array}[c]{cc}H'-E_{F}\hat{1}_{8\times8} & \Delta\hat{1}_{8\times8}\\\Delta^{*} & -(T'H'T'^{-1}-E_{F}\hat{1}_{8\times8})\end{array} \right)
\left(\begin{array}[c]{cc}\Psi^{e}\\\Psi^{h}\end{array}\right)
\end{equation*}
\begin{equation*}
=\varepsilon\left(\begin{array}[c]{cc}\Psi^{e}\\\Psi^{h}\end{array} \right).\tag{A5}
\end{equation*}
Here, $T'$ is the time-reversal operator.
When the spin degrees of freedom are also considered, the time-reversal operator reads
\begin{equation*}
T'=(\sigma_{z}\otimes\tau_{x}\otimes{-iS_{y}})C,\tag{A6}
\end{equation*}
$S_{z}$ and $S_{y}$ are Pauli matrices acting on real spin space and $-iS_{y}$ reverses 
real spin directions.
It can be shown that $T'H'T'^{-1}=H\otimes\hat{1}_{2\times2}-h\hat{1}_{4\times4}\otimes{S_{z}}$ .
The solution of Eq. (A5) must satisfy the condition: $\Psi^{h}=T\Psi^{e}$.
Therefore, by writing $\Psi^{e}=(u^{+}_{\uparrow}\mbox{ }u^{-}_{\uparrow}\mbox{ }u^{+}_{\downarrow}\mbox{ }u^{-}_{\downarrow})$
and  $\Psi^{h}=(v^{+}_{\uparrow}\mbox{ }v^{-}_{\uparrow}\mbox{ }v^{+}_{\downarrow}\mbox{ }v^{-}_{\downarrow})$,
we find that the Bogolubov-de Gennes equation can be decomposed into four decoupled sets
\begin{equation*}
\left(\begin{array}[c]{cc}
        H_{\pm}-E_{F}\hat{1}_{2\times2}-\rho_{\sigma}h\hat{1}_{2\times2} & \Delta\hat{1}_{2\times2}\\
        \Delta^\ast\hat{1}_{2\times2} & -(H_{\pm}-E_{F}\hat{1}_{2\times2}+\rho_{\sigma}h\hat{1}_{2\times2})
        \end{array}\right)
\end{equation*}
\begin{equation*}
\times\left(\begin{array}[c]{cc}
        u_{\sigma}^{\pm}\\\nu_{\bar{\sigma}}^{\mp}
        \end{array}\right)
=\varepsilon\left(\begin{array}[c]{cc}
        u_{\sigma}^{\pm}\\\nu_{\bar{\sigma}}^{\mp}
        \end{array}\right). \tag{A7}
\end{equation*}
This is the Eq. (3) in Sec. II A.

\section*{APPENDIX B: CALCULATION OF THE ANDREEV AND NORMAL REFLECTION AMPLITUDES}
By solving the DBdG equation with ferromagnetism [Eq. (3)], we find the wave functions in each region of the
model FIS junction shown in Fig. 1.
In region F, the wave functions are given by
\begin{equation*}
\psi^{\pm}_{e\sigma}=(1, {\pm}e^{{\pm}{\it i}\alpha_{\sigma}}, 0, 0)^{T}e^{{\pm}{\it i}p_{e\sigma}x}e^{{\it i}qy},
\end{equation*}
\begin{align}
\psi^{\pm}_{h\bar{\sigma}}=(0, 0, 1, {\mp}e^{{\pm}{\it i}\alpha'_{\bar{\sigma}}})^{T}e^{{\pm}{\it i}p_{h\bar{\sigma}}x}e^{{\it i}qy}, \tag{A8}
\end{align}
with
\begin{equation*}
\sin(\alpha_{\sigma})=\frac{\hbar\upsilon_{F}q}{(\varepsilon+E_F+\rho_{\sigma}E_{ex})}
\end{equation*}
\begin{equation*}
\sin(\alpha'_{\bar{\sigma}})=\frac{\hbar\upsilon_{F}q}{(\varepsilon-E_F+\rho_{\sigma}E_{ex})},
\end{equation*}
\begin{equation*}
p_{e\sigma}=\frac{(\varepsilon+E_F+\rho_{\sigma}E_{ex})}{\hbar\upsilon_{F}}\cos(\alpha_{\sigma}),
\end{equation*}
\begin{align}
p_{h{\bar{\sigma}}}=\frac{(\varepsilon-E_F+\rho_{\sigma}E_{ex})}{\hbar\upsilon_{F}}\cos(\alpha'_{\bar{\sigma}}). \tag{A8}
\end{align}
Here, $q$ is the transverse wave vector of the particle and we have assumed that all transverse wave vectors are equal.
$\varepsilon$ is the excitation energy and ${\pm}$ denote the traveling directions of the particles.
In other words, $\psi^{+}_{e\sigma}$ and $\psi^{+}_{h\bar{\sigma}}$ travel the $+x$ direction,
while $\psi^{-}_{e\sigma}$ and $\psi^{-}_{h\bar{\sigma}}$
travel the $-x$ direction.
$\alpha_\sigma$ is the incident angle of the electron, while $\alpha'_{\bar\sigma}$
is the Andreev reflection angle of the hole.

In order to obtain the wave functions in the insulating region, we simply replace $E_F$ with $E_F-V_0$ and let $E_{ex}=0$ in Eq. (A1) and Eq. (A2).
Therefore, in region I, the wave functions are given by
\begin{equation*}
\tilde{\psi}_{e}^{\pm}=(1, {\pm}e^{{\pm}{\it i}\theta}, 0, 0)^{T}e^{{\pm}{\it i}\tilde{p}_{e}x}e^{{\it i}qy},
\end{equation*}
\begin{align}
\tilde{\psi}_{h}^{\pm}=(0, 0, 1, {\mp}e^{{\pm}{\it i}\theta'})^{T}e^{{\pm}{\it i}\tilde{p}_{h}x}e^{{\it i}qy}, \tag{A9}
\end{align}
with
\begin{equation*}
\sin(\theta)=\frac{\hbar\upsilon_{F}q}{(\varepsilon+E_F-V_0)},
\end{equation*}
\begin{equation*}
\sin(\theta')=\frac{\hbar\upsilon_{F}q}{(\varepsilon-E_F+V_0)},
\end{equation*}
\begin{equation*}
\tilde{p}_{e}=\frac{(\varepsilon+E_F-V_0)}{\hbar\upsilon_{F}}\cos(\theta),
\end{equation*}
\begin{align}
\tilde{p}_{h}=\frac{(\varepsilon-E_F+V_0)}{\hbar\upsilon_{F}}\cos(\theta'). \tag{A10}
\end{align}
The indices $\sigma$ and $\bar{\sigma}$ disappear because our insulator is nonmagnetic.
$\theta$ ($\theta'$) denotes the incident angle of the electron (hole) in region I.
The meanings of the other symbols are the same as that in region F.

In region S, the wave functions are given by
\begin{align}
\psi_s^{\pm}=(e^{{\pm}{\it i}\beta}, {\pm}e^{{\pm}{\it i}(\gamma+\beta)}, e^{-{\it i}\phi}, {\pm}e^{{\it i}({\pm}\gamma-\phi)})^{T}e^{{\pm}{\it i}p_{s}x-{\kappa}x}e^{{\it i}qy}, \tag{A11}
\end{align}
with
\begin{equation*}
\beta =\left\{\begin{array}{ll}
   \cos^{-1}(\varepsilon/\Delta_0) & \mbox {if}\quad \varepsilon<\Delta_0\\
   -{\it i}\cosh^{-1}(\varepsilon/\Delta_0) & \mbox {if}\quad \varepsilon>\Delta_0
  \end{array} \right.
\end{equation*}
\begin{equation*}
      \sin(\gamma)=\frac{\hbar\upsilon_{F}q}{E_F+U_0},
\end{equation*}
\begin{equation*}
      p_s= \frac{(E_F+U_0)}{\hbar\upsilon_{F}}\cos(\gamma),
\end{equation*}
\begin{align}
      {\kappa}=\frac{(E_F+U_0)\Delta_0}{\hbar^2\upsilon_{F}^{2}p_s}\sin(\beta). \tag{A12}
\end{align}
Both $\psi_s^{+}$ and $\psi_s^{-}$ travel in the $+x$ direction.
$\psi_s^{+}$ represents the state traveling in the same direction as its wave vector
while $\psi_s^{-}$ means the state traveling in the direction opposite to its wave vector.
$\gamma$ is the angle of incidence of the quasiparticles in region S.

Then, the overall wave functions in region F, region I, and region S read
\begin{equation*}
\Psi_{F\sigma}=\psi^{+}_{e\sigma}+a_{\sigma}\psi^{-}_{h\bar{\sigma}}+b_{\sigma}\psi^{-}_{e\sigma},
\end{equation*}
\begin{equation*}
\Psi_{I}=m_{1}\tilde{\psi}_{e}^{+}+m_2\tilde{\psi}_{e}^{-}+m_{3}\tilde{\psi}_{h}^{+}+m_{4}\tilde{\psi}_{h}^{-},
\end{equation*}
\begin{align}
\Psi_{S}=t_{1}\psi_{s}^{+}+t_{2}\psi_{s}^{-}. \tag{A13}
\end{align}
Note that $a_{\sigma}$ and $b_{\sigma}$ are the amplitudes of the Andreev reflection and
normal reflection, respectively, as spin-$\sigma$ electrons incident.
They can be used to calculate the tunneling conductance and we can obtain them by applying the boundary conditions,
\begin{align}
\Psi_{F\sigma}|_{x=-d}= \Psi_{I}|_{x=-d},\quad\Psi_{I}|_{x=0}=\Psi_{S}|_{x=0}. \tag{A14}
\end{align}

In the thin-barrier limit, we let $d\rightarrow0$ and $V_0\rightarrow\infty$ and introduce the finite barrier strength $\chi=V_0d/\hbar\upsilon_F$
such that $\theta\rightarrow0$, $\theta'\rightarrow0$, $\tilde{k}_ed\rightarrow-\chi$, and $\tilde{k}_ed\rightarrow\chi$.
Then, $a_{\sigma}$ and $b_{\sigma}$ can be written as

\begin{equation*}
a_{\sigma}=\frac{2\cos(\alpha_{\sigma})\cos(\gamma)}{X_{\sigma}}e^{-{\it i}{2\chi}}e^{-{\it i}\phi},
\end{equation*}
\begin{align}
b_{\sigma}=\frac{2\cos(\alpha_{\sigma})X_{1\sigma}-X_{\sigma}}{X_\sigma}e^{-{\it i}2\chi}. \tag{A15}
\end{align}
Here, $X_{\sigma}$, $X_{1\sigma}$ are given by
\begin{equation*}
X_{\sigma}=X_{1\sigma}e^{-{\it i}\alpha_{\sigma}}+X_{2\sigma},
\end{equation*}
\begin{equation*}
X_{1\sigma}=[\cos(\beta)\cos(\gamma)+\sin(\beta)\sin(\gamma)\cos(2\chi)]
\end{equation*}
\begin{equation*}
+{\it i}[\sin(\beta)-\sin(\beta)\sin(\gamma)\sin(2\chi)]e^{-{\it i}{\alpha'_{\bar{\sigma}}}},
\end{equation*}
\begin{equation*}
X_{2\sigma}=[\cos(\beta)\cos(\gamma)-\sin(\beta)\sin(\gamma)\cos(2\chi)]e^{-{\it i}{\alpha'_{\bar{\sigma}}}}
\end{equation*}
\begin{align}
+{\it i}[\sin(\beta)+\sin(\beta)\sin(\gamma)\sin(2\chi)]. \tag{A16}
\end{align}

\section*{APPENDIX C: CALCULATION OF THE TUNNELING SPIN POLARIZATION}
In order to obtain the tunneling spin polarization, we must calculate the tunneling
spin current for the incident spin-$\sigma$ electron first.
For a {\sl s}-wave superconductor, a cooper pair is composed of two electrons with opposite spin directions.
The Andreev reflected hole has the same spin direction $\sigma$ as the incident electron
but travels along the $-x$ direction opposite to the incident electron, as decribed in Sec. II. A.
The motion of Andreev reflected hole with spin-$\sigma$ is equivalent to
that an electron with $\bar\sigma$ tunnels through the F-I-S junction interface.
Therefore, the Andreev reflected hole carries the spin-$\bar{\sigma}$ current ($I_{\bar\sigma}$)
while the incident and normal reflected electron carries the spin-${\sigma}$ current ($I_{\sigma}$).
Therefore, the difference and sum of spin current for the incident spin-$\sigma$
electron are given, respectively, by
\begin{align*}
(I_{\uparrow}-&I_{\downarrow})_{\sigma}=\frac{\rho_{\sigma}}{e}\int^{eV}_{0}d\varepsilon\int^{\alpha_{\sigma}^c(\varepsilon)}_{0}
d\alpha_\sigma{G_{\sigma}(\varepsilon)}\cos(\alpha_\sigma)\times\\
&(1-|a_\sigma(\varepsilon,\alpha_\sigma)|^2\frac{\cos(\alpha'_{\bar{\sigma}})}{\cos(\alpha_{\sigma})}
-|b_\sigma(\varepsilon,\alpha_\sigma)|^2),
\end{align*}
\begin{align*}
(I_{\uparrow}+&I_{\downarrow})_{\sigma}=\frac{1}{e}\int^{eV}_{0}d\varepsilon\int^{\alpha_{\sigma}^c(\varepsilon)}_{0}
d\alpha_\sigma{G_{\sigma}(\varepsilon)}\cos(\alpha_\sigma)\times\\
&(1+|a_\sigma(\varepsilon,\alpha_\sigma)|^2\frac{\cos(\alpha'_{\bar{\sigma}})}{\cos(\alpha_{\sigma})}
-|b_\sigma(\varepsilon,\alpha_\sigma)|^2). \tag{A17}
\end{align*}
Then, summing up the contributions of the incident spin-up and spin-down electrons, we obtain the tunneling spin polarization
\begin{align*}
P_{T}=\frac{(I_{\uparrow}-I_{\downarrow})}{(I_{\uparrow}+I_{\downarrow})}
=\frac{\sum_{\sigma}(I_{\uparrow}-I_{\downarrow})_{\sigma}}{\sum_{\sigma}(I_{\uparrow}+I_{\downarrow})_{\sigma}}. \tag{A18}
\end{align*}

\end{document}